\documentclass{article}

\title{Coulomb scattering of the Dirac fermions on de Sitter expanding
universe}
\author{Cosmin Crucean\\
{\normalsize West University of Timi\c soara, V. Parvan}\\
\normalsize Avenue 4 RO-1900 Timi\c soara, Romania}
\begin{document}
\maketitle

\begin{abstract}
The lowest order contribution of the amplitude of the Dirac -
Coulomb scattering in de Sitter spacetime is calculated assuming
that the initial and final states of the Dirac field are described
by exact solutions of the free Dirac equation on de Sitter
spacetime with a given momentum and helicity. One studies the
difficulties that arises when one passes from the amplitude to
cross section.
\end{abstract}

\section{Introduction}
\par
It is known that many difficult problems arises when one tries to
combine the quantum theory of fields with general relativity. One
of the main problems is related to the physical interpretation of
the one-particle quantum modes that may indicate how to quantize
the fields. The curved speacetime have specific symmetries which,
in general, differ from that of the Minkowski spacetime. For this
reason, here the symmetries generating conserved quantities have
to be treated with specific methods. In other respects, it is
known that the form of the fields equations and implicitly their
solutions on curved spacetime are strongly dependent on the tetrad
gauge and local chart in which one works. It is important to point
out that in the latest years important steps in developing a
quantum field theory on de Sitter spacetime was made by finding
analytical solutions for the Dirac equation in moving or static
local charts suitable for separation of variables [2], [8], [9],
[10].
 \par
 In the present the majority of investigations dedicated to Q.E.D
 on curved spacetime do not take in considerations
 scattering processes. Our aim in this paper is to calculate one
 scatter amplitude for Dirac field in curved spacetime and to analyze
the physical
 consequences that emerge from this calculation.
 Differences with respect to the Minkowski case are that the modulus
 of the momentum is not conserved and a tendency for helicity
 conservation. Also in the limit of a vanishing expansion rate of
 the space the Minkowski amplitude will not be recovered. Finally, we
 obtain that the cross section is a sum of two contributions, that
corresponds to linear and nonlinear amplitude of scattering.
 \par
 In section 2 we derive the formula for the scattering
 amplitude. In section 3 we discuss some of proprieties of
 scatter amplitude and we will examine the limit cases of a
 small/large expansion rate of the space compared with the mass of
 particle, in section 4 we calculate the cross section.
 Our conclusions are summarized in section 5 pointing
 out a series of aspects which remain to be clarified about this
 subject. The results are presented in natural units $\hbar=c=1$.

 \section{The scattering amplitude}

  We start with the exact solutions
  of the free Dirac equation in the de Sitter spacetime written in
  [2]. Let us write the de Sitter line element [1],
  \begin{equation}
  ds^{2}=dt^{2}-e^{2\omega t}d\vec{x}^{2},
  \end{equation}
  where $\omega$ is the expansion factor and $\omega>0$.
  Now we know that defining a spinor field on curved spacetime requires
  one to use the tetrad fields [1] $e_{\hat{\mu}}(x)$ and
$\hat{e}^{\hat{\mu}}(x)$,
  fixing the local frames and corresponding coframes which are labelled
by the local
  indices $\hat{\mu},\hat{\nu},...=0,1,2,3$. The form of the
  line element allows one to chose the simple Cartesian gauge with
  the non-vanishing tetrad components:
  \begin{equation}
  e^{0}_{\hat{0}}=e^{-\omega t};
\quad e^{i}_{\hat{j}}=\delta^{i}_{j}e^{-\omega t},
  \end{equation}
  so that $e_{\hat{\mu}}=e^{\nu}_{\hat{\mu}}e_{\nu}$ and
  have the orthonormalization properties
  $e_{\hat{\mu}}e_{\hat{\nu}}=\eta_{\hat{\mu}\hat{\nu}},\\
  \hat{e}^{\hat{\mu}}e_{\hat{\nu}}=\delta^{\hat{\mu}}_{\hat{\nu}}$
  with respect to the Minkowski metric $\eta={\rm diag}(1,-1,-1,-1)$.
  \par
Now let us introduce normalized helicity spinors for an arbitrary
vector $\vec{p}$ by notation: $\xi_{\lambda}(\vec{p})$,
\begin{equation}
\vec{\sigma}\vec{p}\xi_{\lambda}(\vec{p})=2p\lambda\xi_{\lambda}(\vec{p}),
\end{equation}
with $\lambda=\pm 1/2$ and where $\vec{\sigma}$ are the Pauli
matrices and $p=\mid\vec{p}\mid$. For witting the solutions of
Dirac equation on de Sitter spacetime we set:
\begin{equation}
k=\frac{m}{\omega},\quad \nu_{\pm}=\frac{1}{2}\pm ik
\end{equation}
Then the positive frequency modes of momentum $\vec{p}$ and
helicity $\lambda$ that were constructed in [2] using the gamma
matrices in Dirac representation (with diagonal $\gamma^0$) are:
\begin{equation}
U_{\vec{p},\lambda}(t,\vec{x})=\frac{\sqrt{\pi
p/\omega}}{(2\pi)^{3/2}}\left (\begin{array}{c} \frac{1}{2}e^{\pi
k/2}H^{(1)}_{\nu_{-}}(\frac{p}{\omega} e^{-\omega
t})\xi_{\lambda}(\vec{p})\\
\lambda e^{-\pi k/2}H^{(1)}_{\nu_{+}}(\frac{p}{\omega} e^{-\omega
t})\xi_{\lambda}(\vec{p})
\end{array}\right)e^{i\vec{p}\vec{x}-2\omega t},
\end{equation}
where $H^{(1)}_{\nu}(z)$ is the Hankel function of first kind.
\par
Since the charge conjugation  in a curved background is point
independent [7], as in Minkowski case, the negative frequency
modes can be obtained using the charge conjugation,
\begin{equation}
U_{\vec{p},\lambda}(x)\rightarrow
V_{\vec{p},\lambda}(x)=i\gamma^{2}\gamma^{0}(\bar{U}_{\vec{p},\lambda}(x))^{T}\,.
\end{equation}
Thus we can restrict ourselves to analyze only the positive
frequency modes. These spinors satisfy the orthonormalization
relations[2]:
\begin{eqnarray}
\int d^{3}x
(-g)^{1/2}\bar{U}_{\vec{p},\lambda}(x)\gamma^{0}U_{\vec{p^{\prime}},\lambda^{\prime}}(x)=\\
\nonumber\int d^{3}x
(-g)^{1/2}\bar{V}_{\vec{p},\lambda}(x)\gamma^{0}V_{\vec{p^{\prime}},\lambda^{\prime}}(x)=
\delta_{\lambda\lambda^{\prime}}\delta^{3}(\vec{p}-\vec{p^{\prime}})\\
\nonumber\int d^{3}x
(-g)^{1/2}\bar{U}_{\vec{p},\lambda}(x)\gamma^{0}V_{\vec{p^{\prime}},\lambda^{\prime}}(x)=0,
\end{eqnarray}
where the integration extends on an arbitrary hypersurface
$t=const$ and $(-g)^{1/2}=e^{3\omega t}$. They represent a
complete system of solutions in the sense that[2]
\begin{equation}
\int d^{3}p
\sum_{\lambda}\left[U_{\vec{p},\lambda}(t,\vec{x})U^{+}_{\vec{p},\lambda}(t,\vec{x^{\prime}})+
V_{\vec{p},\lambda}(t,\vec{x})V^{+}_{\vec{p},\lambda}(t,\vec{x^{\prime}})\right]=e^{-3
\omega t}\delta^{3}(\vec{x}-\vec{x^{\prime}})\,.
\end{equation}
\par
In our calculation we need to find the form of the Coulomb
potential on de Sitter spacetime which depends on metric. Here we
can exploit the conformal invariance of the Maxwell equations
since the de Sitter metric is conformal with the Minkowski one. We
can write $\frac{Ze}{|\vec{x}|}$ for the Coulomb field in
Minkowski spacetime. Then we find  the corresponding de Sitter
potential:
\begin{equation}
A^{\hat{0}}(x)=\frac{Ze}{|\vec{x}|} e^{-\omega t},\quad
A^{\hat{j}}(x)=0,
\end{equation}
where the hated indices indicate label the components in the local
Minkowski frames. We also observe that Eq. (9) is just the
expression from flat space with distances dilated/contracted by
the factor $e^{-\omega t}$.
\par
Having the above elements in our mind we can proceed to develop
the theory for the scattering amplitude on de Sitter spacetime
which can be reproduced from that in the Minkowski space. The
necessary requirements for develop the scattering theory is the
global hyperbolicity of the space and having a complete set of
solutions of the free equation for incident field and scattered
field (Born approximation) with the distinction between positive
and negative frequencies. It is also important to specify that in
our analysis both cases are fulfilled.
\par
Let $\psi_{i}(x)$ and $\psi_{f}(x)$ be the waves freely
propagating in the {\em in} and {\em out} sectors, and we assume
that they are both of positive frequency. In direct analogy with
the Minkowski [3], [4] theory we can define the lowest order
contribution in the scattering amplitude as follows:
\begin{equation}
A_{i\rightarrow f}=-ie \int d^{4}x
\left[-g(x)\right]^{1/2}\bar\psi_{f}(x)\gamma_{\mu}A^{\hat{\mu}}(x)\psi_{i}(x),
\end{equation}
where $e$ is the unit charge of the field. Our intention is to
calculate the amplitude of Coulomb scattering
 for the external field (9) and for initial and final states of
 the form
 \begin{equation}
\psi_{i}(x)=U_{\vec{p_{i}},\lambda_{i}}(x),
\psi_{f}(x)=U_{\vec{p_{f}},\lambda_{f}}(x).
 \end{equation}
Now replacing in Eq. (10) the explicit form of the Dirac spinors
and the form of the Coulomb potential we observe two remarkable
properties. First is that the dependence of $t, \vec{x}$ allows us
to split the four dimensional integral into a pure spatial
integral and a temporal one. The second property is that the pure
spatial integral have exactly the same form as in the Minkowski
space.
\begin{equation}
\int d^{3}x
\frac{e^{i(\vec{p_{i}}-\vec{p_{f}})\vec{x}}}{|\vec{x}|}=\frac{4
\pi}{|\vec{p_{f}}-\vec{p_{i}}|^{2}}
\end{equation}
It is clear that the not so simple part is the temporal integral
which contains the influence of the gravitational field via the
expansion parameter,
\begin{eqnarray}
&&
 A_{i\rightarrow f}=-i\alpha Z
\frac{\sqrt{p_{i}p_{f}}}{8\pi|\vec{p_{f}}-\vec{p_{i}}|^{2}}\xi^{+}_{\lambda_{f}}(\vec{p_{f}})\xi_{\lambda_{i}}(\vec{p_{i}})
\left[e^{\pi k}\int_0^{\infty} dz
zH^{(2)}_{\nu_{+}}(p_{f}z)H^{(1)}_{\nu_{-}}(p_{i}z)\right.\nonumber\\
&&\left.+sgn(\lambda_{f}\lambda_{i})e^{-\pi k}\int_0^{\infty} dz
zH^{(2)}_{\nu_{-}}(p_{f}z)H^{(1)}_{\nu_{+}}(p_{i}z)\right],
\end{eqnarray}
where we pass to a new variable of integration
\begin{equation}
z=\frac{e^{-\omega t}}{\omega}.
\end{equation}
Note that the integration limits in (13) corresponds to
$t=\pm\infty$ , we assume that the interaction extends into the
past and future.
\par
For simplification of our notation we introduce the following
quantities:
\begin{equation}
A_{k}(f,i)=A^{+}_{k}(p_{f}p_{i})+sgn(\lambda_{f}\lambda_{i})A^{-}_{k}(p_{f}p_{i}),
\end{equation}
with
\begin{equation}
A^{\pm}_{k}(p_{f}p_{i})=\frac{\sqrt{p_{i}p_{f}}}{2}e^{\pm\pi
k}\int_0^{\infty} dz
zH^{(2)}_{\nu_{\pm}}(p_{f}z)H^{(1)}_{\nu_{\mp}}(p_{i}z).
\end{equation}
Putting all the above notations together we can write for the
scattering amplitude:
\begin{equation}
A_{i\rightarrow f}=-\frac{i}{4\pi}\frac{\alpha
Z}{|\vec{p_{f}}-\vec{p_{i}}|^{2}}A_{k}(f,i)\xi^{+}_{\lambda_{f}}(\vec{p_{f}})\xi_{\lambda_{i}}(\vec{p_{i}}),
\end{equation}
with $\alpha=e^2$ and $\xi_{f}, \xi_{i}$ stand for the initial and
final helicity two spinors.
\par
The evaluation of the integrals is discussed in Appendix, here we
give the final result in terms of Dirac delta-function,
hypergeometric functions, Euler Beta functions and unit step
function
\begin{eqnarray}
A_{k}(f,i)=\delta(p_{f}-p_{i})+\theta(p_{i}-p_{f})\frac{1}{p_{i}}
f_{k}\left(\frac{p_{f}}{p_{i}}\right)+\theta(p_{f}-p_{i})\frac{1}{p_{f}}f^{*}_{k}\left(\frac{p_{i}}{p_{f}}\right)+\nonumber\\
sgn(\lambda_{f}\lambda_{i})\left[\delta(p_{f}-p_{i})+\theta(p_{i}-p_{f})\frac{1}{p_{i}}
f_{-k}\left(\frac{p_{f}}{p_{i}}\right)+\theta(p_{f}-p_{i})\frac{1}{p_{f}}f^{*}_{-k}\left(\frac{p_{i}}{p_{f}}\right)\right].
\end{eqnarray}
We introduce the following notation for simplify our formula:
\begin{eqnarray}
f_{k}\left(\chi \right)=i\left(\chi \right)^{-i k}\frac{e^{\pi
k}}{cosh(\pi k)}\frac{_{2}F_{1}\left(\frac{1}{2},1-i
k;\frac{1}{2}-i
k;\chi^{2}\right)}{B\left(\frac{1}{2},\frac{1}{2}-ik\right)}\\
\nonumber-i\left(\chi\right)^{1+ik}\frac{e^{-\pi k}}{cosh(\pi
k)}\frac{_{2}F_{1}\left(\frac{3}{2},1+i k;\frac{3}{2}+i
k;\chi^{2}\right)}{B\left(-\frac{1}{2},\frac{3}{2}+ik\right)},
\end{eqnarray}
where $\chi=\frac{p_{f}}{p_{i}}$ or $\frac{p_{i}}{p_{f}}$ and
$f_{-k}\left(\chi\right)$ is obtain when $k\rightarrow -k$ in
(19).
\par
The above formulas is our result and in remaining paper we will
explore some of their physical consequences. Before that we must
state that the argument $\chi$ in (19) must be considered in
interval $0\leq\chi<1$ (note the argument in the $\theta$
functions). This is the domain of convergence of the
hypergeometric functions, because in the limit $\chi\rightarrow1$
the $_{2}F_{1}(a,b,c,\chi^{2})$ functions diverge. The above
observations are crucial when we compare our expression of
scattering amplitude with the well known expression of scattering
amplitude from Minkowski spacetime. In our case we see that in
(19) appear terms that demand $p_{f}\neq p_{i}$ and from this we
conclude that there exist a nonvanishing probability for a
scattering process where the law of conservation for total
momentum is lost.

\section{Limit cases}

It is clear that the influence of the de Sitter geometry in the
amplitude is contained in the $A_{k}(f,i)$ factor which was
obtained after integration with respect to time variable. This
quantity encode via the dependence of the $k$ parameter the effect
of the expansion of the space on the scattering process.
\par
We start our analysis having in mind that for $\omega=0$ we obtain
from de Sitter metric the Minkowski one and this corresponds to
$k= \infty$. We see that in our expression just the terms with
Dirac delta-functions will not vanish. Also for opposite
helicities $A_{\infty}(f,i)=0$ and for equal helicities we will
obtain:
\begin{equation}
A_{\infty}(f,i)=2\delta_{\lambda_{f}\lambda_{i}}\delta(p_{f}-p_{i}).
\end{equation}
The correspondent factor from Minkowski case for equal helicities
is:
\begin{equation}
 A_{Minkowski}(f,i)=4\delta(E_{f}-E_{i}),
\end{equation}
where the notations are : $E_{f},E_{i}$ are the final and initial
Minkowski energies and $E^{2}=p^{2}+m^{2}$. Now using the
properties of distributions and the fact that helicity
conservation increase in the ultra relativistic limit ($p\simeq
E\gg m$) we obtain the amplitude ratio deSitter/Minkowski:
\begin{equation}
\frac{2\delta(p_{f}-p_{i})}{4\delta(E_{f}-E_{i})}\simeq\frac{1}{2}.
\end{equation}
The fact that we don't recover the Minkowski amplitude can be
explained as follows. As we point out in the previous section the
limits of integration in (13) for the time variable extends into
the past and future $t\rightarrow\pm\infty$.   Observing that for
$t\rightarrow\infty$ the integrand vanishes as $e^{-\omega t}$,
while for $t\rightarrow-\infty$ become divergent. In this case we
can use the asymptotic formulas for Hankel functions
\begin{equation}
H^{(1,2)}_{\nu}(z)=\left(\frac{2}{\pi z}\right)^{1/2}e^{\pm
i(z-\nu\pi/2-\pi/4)}, \quad z\rightarrow\infty.
\end{equation}
Passing to conformal time $t_{c}=-z$ and using (23) the
contributions to the amplitude in (16) will be :
\begin{equation}
\frac{1}{\pi}\int dt_{c}e^{i(p_{f}-p_{i})t_{c}} , \quad
t_{c}\rightarrow -\infty.
\end{equation}
The integral (24) produces the distributional term, integrating
over an infinite interval. Observing that in (24) appears only one
semi-infinite axis the explanation for the 1/2 factor is
immediate. To summarize in deSitter case the distributional terms
will be produced by the evolution in the infinite past when the
space was contracted. This contraction of the space in the
infinite past is responsible for the 1/2 factor when we pass to
Minkowski limit.
\par
We will make a few comments about nonconservation  of total
momentum. The difference between (21) and the result in the
previous section (18) is that the last one does not vanish for
$p_{f}\neq p_{i}$. As we point out in the previous section the
total momentum is not conserved in the scattering process on de
Sitter spacetime. We know that in Minkowski case the conservation
of the momentum modulus is a direct consequence of the
conservation of total energy which in turns is a consequence of
the translational invariance with respect to time, which is lost
in de Sitter space.
\par
Now let us take a closer analysis to helicities from which we
deduce a interesting properties. First we observe that the delta
terms $\delta(p_{f}-p_{i})$ in (18), will add for equal helicities
and vanish for opposite ones. Now if we admit that we deal with a
process such that modulus of the final momentum is closer to the
modulus of the initial momentum $p_{f}\sim p_{i}$ the probability
of transition between different helicities states will be smaller
that that between identical helicity states. From that we can
conclude that on de Sitter space is manifesting a tendency of
conservation for helicity.
\par
Let us consider now the opposite case $k=0$ which corresponds to a
very large expansion rate of the space compared with mass of
particle. The $f_{\pm k}(\chi)$ functions simplify in this case,
observing that for null $k$ the second and the third argument in
hypergeometric functions became the same and using
$_{2}F_{1}(a,b;b;z)=(1-z)^{-a}$, we find
\begin{equation}
f_{(k=0)}\left(\chi\right)=\frac{1}{\pi}\frac{i}{1-\chi}.
\end{equation}
Introducing (25) in (18) we obtain a pair of terms which summed
make the unit step functions unnecessary. Giving attention to the
delta contributions, one obtain for identical initial and final
helicities:
\begin{equation}
A_{0}(f,i)=2\delta(p_{f}-p_{i})+\frac{4}{\pi}\frac{i}{p_{i}-p_{f}}.
\end{equation}
For opposite helicities the result is identically null, since for
$k=0$ it is easy to see that $A^{\pm}_{k}(p_{f},p_{i})$ exactly
cancel.
\par
An important observation is that (26) can be obtained if we put
$k=0$ in Hankel functions obtaining,
\begin{equation}
H^{(1,2)}_{1/2}(z)=\mp i\left(\frac{2}{\pi z}\right)^{1/2}e^{\pm
iz},
\end{equation}
in which case the integrals in (16) can be easy evaluated.
\par
Now we know that the Dirac theory is conformally invariant for a
null fermion mass, and that our space is conformal with the
Minkowski one. The case $k=0$ is equivalent with a vanishing mass
$m=0$ and the explanation is immediate. Also note that the
massless limit is  consistent with the identically null amplitude
in the case when  helicity is not conserved.
\par
In the above limit cases we observe that the amplitude will vanish
for opposite helicities and the conclusion is immediate: in bought
cases the total angular momentum is conserved. As we point out
above this tendency is also manifested in the general case.

\section{Calculation of cross section}

We know from the scattering theory that the parameter measured by
experiment in scattering processes is the cross section. In our
case the situation is complicated and we must reconsider our
vision about cross section. The first observation is that our
amplitude of scattering contain two main contributions one linear
part (terms with delta Dirac function) and one non-linear part. In
this situation it is natural to define in general the differential
cross section as a sum of a linear contribution and a non-linear
one. Also it is important to interpret our non-linear amplitude at
square as a probability of scatter when the modulus of momentum is
not conserved. The definition of linear cross section for Coulomb
scatter can be reproduced as in Minkowski spacetime
\begin{equation}
d\sigma_{l}=\frac{1}{2}\sum_{\lambda_{i}\lambda_{f}}\frac{1}{J}\frac{dP_{l}}{dt},
\end{equation}
where the factor $\frac{1}{2}$ transform one sum in mediation,
 $P_{l}$ is the linear probability of transition and $J$
is the incident flux in unit of time. When modulus of momentum is
conserved just the linear probability of transition will give
contribution to cross section.
\par
For the non-linear differential cross section for Coulomb scatter
we will use the following definition
\begin{equation}
d\sigma_{nl}=\frac{1}{2}\sum_{\lambda_{i}\lambda_{f}}\frac{1}{J}dP_{nl},
\end{equation}
where $P_{nl}$ is the non-linear probability of transition. In the
case when modulus of momentum is not conserved just the non-linear
probability of transition will give contribution to cross section.
\par
For our further calculations we evaluate the incident flux in unit
of time using the formula of Dirac current in local frames:
\begin{equation}
J^{\hat\mu}=e^{\hat\mu}_{\nu}\bar
U_{\vec{p_{i}},\lambda_{i}}(x)\gamma^{\nu}U_{\vec{p_{i}},\lambda_{i}}(x).
\end{equation}
The spatial component of Dirac current can be defined in our case
as follows:
\begin{eqnarray}
&&
 J=\frac{e^{\omega t}}{\mid \vec{p_{i}}\mid }\bar
U_{\vec{p_{i}},\lambda_{i}}(x)(\vec{p_{i}}\vec{\gamma}
)U_{\vec{p_{i}},\lambda_{i}}(x)= \frac{\pi p_{i} e^{-3\omega
t}}{(2\pi)^{3}4\omega}\left[H^{(2)}_{\nu_{-}}(\frac{p_{i}}{\omega}
e^{-\omega t})H^{(1)}_{\nu_{-}}(\frac{p_{i}}{\omega} e^{-\omega
t})\right.\nonumber\\
&&\left.+H^{(2)}_{\nu_{+}}(\frac{p_{i}}{\omega} e^{-\omega
t})H^{(1)}_{\nu_{+}}(\frac{p_{i}}{\omega} e^{-\omega t})\right].
\end{eqnarray}
The spatial components of the four vector current of particles is
the incident flux in unit of time. We see that this is a quantity
that depends of time which is a unusual and complex problem. In
addition another problem appears, we known that when we pass from
scattering amplitude to cross section an essential role in the
treatment when we deal with finitely extended wave is played by
the energy conservation factor $\delta(E_{f}-E_{i})$ . This factor
assures that the cross section will not depend on the particular
form of wave function of the incident particles. In our case the
cross section calculate with our amplitude will depend on the
specific form of the incident wave which is another unusual
situation. All these aspects are consequences of the lost
invariance with respect to time in de Sitter spacetime.
\par
For calculating the cross section in this situation we will follow
the next physical picture. According to the formalism of the
scattering theory the amplitude obtained here has to be associated
with the following picture: the in states is freely propagated
from the moment $t=0$ back to the infinite past, then evaluated in
presence of the interaction up to infinite future, and again
freely propagated back to $t=0$ where is projected in out states.
In this situation we can evaluate Eq. (31) at $t=0$, thus
obtaining one quantity that is not dependant of time,
\begin{equation}
J=\frac{\pi p_{i}
}{4\omega(2\pi)^{3}}\left[H^{(2)}_{\nu_{-}}(\frac{p_{i}}{\omega}
)H^{(1)}_{\nu_{-}}(\frac{p_{i}}{\omega}
)+H^{(2)}_{\nu_{+}}(\frac{p_{i}}{\omega}
)H^{(1)}_{\nu_{+}}(\frac{p_{i}}{\omega})\right].
\end{equation}
When we deal with particles with given helicity in processes of
scattering we must average upon the helicities of
 incident particles and sum upon the helicities of the emergent
particles. In our case the situation is immediate and we obtain:
\begin{equation}
\frac{1}{2}\sum_{\lambda_{i}\lambda_{f}}\left[\xi^{+}_{\lambda_{f}}(\vec{p_{f}})\xi_{\lambda_{i}}(\vec{p_{i}})\right]^{2}=
2.
\end{equation}
For calculation of linear cross section we need to evaluate
$\frac{dP_{l}}{dt}$ thus obtaining
\begin{eqnarray}
\frac{dP_{l}}{dt}=\frac{(\alpha
Z)^{2}}{8\pi^{3}|\vec{p_{f}}-\vec{p_{i}}|^{4}}\delta(p_{f}-p_{i})d^{3}p_{f}.
\end{eqnarray}
Replacing(34) and (32) in (28) writing
$d^{3}p_{f}=p^{2}_{f}dp_{f}d\Omega_{p_{f}}$
 and integrating with respect to
$p_{f}$ the factor with Dirac delta-function will be eliminated
 and in addition using that $p_{f}=p_{i}=p$ we will obtain for the
linear part:
\begin{equation}
\frac{d\sigma_{l}}{d\Omega}=\frac{(\alpha Z)^{2}\omega}{2\pi
p^{3}\sin^{4}\left(\frac{\theta}{2}\right)}\left[H^{(2)}_{\nu_{-}}(\frac{p}{\omega}
)H^{(1)}_{\nu_{-}}(\frac{p}{\omega}
)+H^{(2)}_{\nu_{+}}(\frac{p}{\omega}
)H^{(1)}_{\nu_{+}}(\frac{p}{\omega})\right]^{-1}.
\end{equation}
\par
We see that our cross section have the usual angular dependence as
the cross section in Minkowski case. In the $k=0$ limit we will
obtain a much simple formula:
\begin{equation}
\left(\frac{d\sigma_{l}}{d\Omega}\right)_{k=0}=\frac{(\alpha
Z)^{2}}{8p^{2}\sin^{4}\left(\frac{\theta}{2}\right)}.
\end{equation}
Now for calculation of cross section in the Minkowski limit
($k=\infty $) we must approximate the Hankel functions
$H^{(1,2)}_{\nu}(z)$ when bought arguments $\nu$ and $z$ have
large values which is one of the most difficult problems. In our
case one of the argument is imaginal and the other real and in the
theory of cylindrical functions are given just the asymptotic
formulas when bought are real or imaginal. Thus this problem of
taking this limit is not solved, our calculations in this
direction shows that this is a very difficult task. For this
reason we can't evolve the cross section in the Minkowski limit.
\par
 In the non-linear case the integration with respect to $p_{f}$
give an expression which is to complicated to be given here. For
this reason we restrict to give an expression for the differential
probability of scatter in this case. Formulas necessary to solve
integrals with respect to $p_{f}$ for the calculation of cross
section is given in Appendix B.
\begin{eqnarray}
&& dP_{nl}=\frac{(\alpha
Z)^{2}}{16\pi^{2}|\vec{p_{f}}-\vec{p_{i}}|^{4}}\left[\theta(p_{i}-p_{f})\frac{1}{p^{2}_{i}}\left[M_{k}\left(\frac{p_{f}}{p_{i}}\right)
+sgn(\lambda_{i}\lambda_{f})B_{k}\left(\frac{p_{f}}{p_{i}}\right)\right]+\right.\nonumber\\
&&\left.\theta(p_{f}-p_{i})\frac{1}{p^{2}_{f}}\left[M_{k}\left(\frac{p_{i}}{p_{f}}\right)+
sgn(\lambda_{i}\lambda_{f})B_{k}\left(\frac{p_{i}}{p_{f}}\right)\right]\right]d^{3}p_{f}
\end{eqnarray}
In the above relation we introduce the following notations:
\begin{eqnarray}
B_{k}\left(\chi\right)&=&f_{k}\left(\chi\right)f^{*}_{-k}\left(\chi\right)+
f^{*}_{k}\left(\chi\right)f_{-k}\left(\chi\right)\,,\nonumber\\
M_{k}\left(\chi\right)&=&\left|f_{k}\left(\chi\right)\right|^{2}+\left|f_{-k}\left(\chi\right)\right|^{2}\,.
\end{eqnarray}
\par
 In the $k=0$ limit we can evolve the non-linear contribution to
cross section making the integral over final momentum and restrict
the limit of integration to $p_{fmax}$ because if we take the
integrals between $0$ and $\infty$ our integral diverge. In fact
this restriction to the limits of integration is suggested by the
form of the scattering amplitude (26) observing that the final
momentum can't exceed one maxim value.
\begin{equation}
\left(\frac{d\sigma_{nl}}{d\Omega}\right)_{k=0}=\frac{16(\alpha
Z)^{2}}{\pi|\vec{p_{f}}-\vec{p_{i}}|^{4}}\frac{[p_{fmax}(2p_{i}-p_{fmax})-2p_{i}(p_{i}-p_{fmax})\ln(\frac{p_{i}}{p_{i}-p_{fmax}})]}
{(p_{i}-p_{fmax})}
\end{equation}
We see that the cross sections calculated with non-linear part of
amplitude have much complicated expressions. In the general case
as we can see from (37) the non-linear cross section will depend
on the parameter of expansion and will have a complicated
expression, which is a sum of expressions of the form (44)(see
Appendix B). For very small values of $\omega$ the spacetime
appears as flat and contributions that contain negative powers of
$k$ can be neglected. In this approximation contributions to
non-linear cross section will come from terms which are
proportional to $e^{2\pi k}$ and $\chi^{2ik}$.

\section{Conclusions}

In this paper we investigated the Coulomb scattering amplitude for
the Dirac field in the expanding de Sitter space. We have complete
ignored complications due to the ambiguity of the particle concept
in a curved spacetime. We considered the initial and final states
of the field as exact solutions of the free Dirac equation in de
Sitter space in the momentum and helicity basis. We also found
that the amplitude depends in an essentially way on the parameter
$k=m/\omega$. For a vanishing $k$ we recover the expected result
due to the conformal invariance of the theory in the massless
case.
\par
Also we found that in the vanishing rate of expansion the
Minkowski amplitude is not recovered. We explain this as a
consequence of the contraction of space in the infinite past.
\par
Now we want to draw attention to some aspects that remaind
untouched in the paper and which could be helpful for further
investigations. A question of principle arises in our paper: how
to define the cross section when this is a quantity which is
dependant of the form of incident wave and in addition the
incident flux is also a dependant of time quantity. Thus when we
calculate the cross section we evolve the incident flux at $t=0$.
This is the only way to calculate the cross section in our
situation and suggest a possible principle when we evolve the
cross section in a spacetime where the translational symmetry with
respect to time is lost. Also we found that the cross section can
be defined in generally as a sum of a linear and a non-linear
part. From our point of view the next important step is to
calculate one scatter amplitude when the projectile particle are
at finite distances from target, because in this situation we may
think to a better connection to experiment. It is also important
to say that an experiment in our situation will depend of the
orientation of the axes of local tetrad.
\par
All these difficult problems that arises in our paper are
consequences of the lost invariance with respect to time in de
Sitter spacetime.
\section{Appendix A:Integrals of Bessel functions}
We will present here the main steps leading to (18). The formulas
that we need are [5],[6]
\begin{eqnarray}
H^{(1)}_{\mu}(z)=\frac{J_{-\mu}(z)-e^{-i\pi\mu}J_{\mu}(z)}{i\sin(\pi\mu)}\nonumber\\
H^{(2)}_{\mu}(z)=\frac{e^{i\pi\mu}J_{\mu}(z)-J_{-\mu}(z)}{i\sin(\pi\mu)}
\end{eqnarray}
and integrals of the type Weber-Schafheitlin [5]
\begin{equation}
\int^{\infty}_{0}dzz^{-s}J_{\mu}(\alpha z)J_{\nu}(\beta z).
\end{equation}
 These integrals can be
solved in the assumptions,
\begin{eqnarray}
\beta>\alpha>0,Re(s)>-1,Re(\mu+\nu-s+1)>0\nonumber\\
\alpha>\beta>0,Re(s)>-1,Re(\mu+\nu-s+1)>0,
\end{eqnarray}
bought cases are fulfilled in our analysis.
 For obtaining (18) we must introduce (40) in (16) and using
Weber-Schafheitlin integrals we will arrive at the desired result.
Now one sees that our integrals in (16) demand $s=-1$. This is
problematic because the integral becomes oscillatory for $z
\rightarrow\infty$. Then a simple way to solve this problem is to
consider an parameter $s$ of the form:
\begin{equation}
s=-1+\epsilon,
\end{equation}
and let in the end $\epsilon \rightarrow 0$. No notably
differences appear when evaluating the integrals in (16) directly
with $s=-1$ and with $\mu,\nu$ arguments following from (4).
\par
Let us take a look to the distributional contribution
$\delta(p_{f}-p_{i})$. This originates in the term
$\delta(\alpha-\beta)$ which is generally present in the integrals
(41) for $s=-1$ (the case of real arguments $\mu=\nu$ is a
familiar one). In our case we must obtain this term for arbitrary
$\mu,\nu$ arguments, because in our case they depend of the
parameter $k$. In watt follows we shall briefly indicate below a
method to obtain it. We start with $s=-1+\epsilon$ parameter like
in (43) and integrate in (41) with respect to $\beta$ over the
interval $\alpha-\epsilon\leq\beta\leq\alpha+\epsilon$. Then the
coefficient of $\delta(\alpha-\beta)$ will be obtained then as the
quantity which is left in the limit $\epsilon\rightarrow 0$.
\section{Appendix B:Integrals with hypergeometric functions}
For calculation of non-linear cross section we solve integrals of
the form:
\begin{eqnarray}
&&
\int_0^{\infty}\theta(p_{i}-p_{f})_{2}F_{1}(b,c;d;\frac{p^{2}_{f}}{p^{2}_{i}})
_{2}F_{1}(e,f;g;\frac{p^{2}_{f}}{p^{2}_{i}})p^{n}_{f}dp_{f}\nonumber\\
&=&\frac{1}{2}\left(-\frac{1}{p_{i}^{2}}\right)^{-n/2}p_{i}^{2}\Gamma(d)\left[\frac{\Gamma(-b+c)\Gamma(-1+b+e-\frac{n}{2})\Gamma(g)\Gamma(-1+b+f-\frac{n}{2})}
{\Gamma(c)\Gamma(-b+d)\Gamma(-1+b+g-\frac{n}{2})\Gamma(e)\Gamma(f)}\right.\nonumber\\
&&\left.\times\Gamma(1-b+\frac{n}{2})_{4}F_{3}(b,1+b-d,-1+b+e-\frac{n}{2},-1+b+f-\frac{n}{2};1+b-c\right.\nonumber\\
&&\left.b-\frac{n}{2},-1+b+g-\frac{n}{2};1)
-\sqrt{-\frac{1}{p^{2}_{i}}}p_{i}\frac{\Gamma(-\frac{1}{2}+b+f-\frac{n}{2})\Gamma(\frac{1}{2}(1-2b+n))}
{\Gamma(c)\Gamma(-b+d)\Gamma(-\frac{1}{2}+b+g-\frac{n}{2})}\right.\nonumber\\
&&\left.\times\frac{\Gamma(-b+c)\Gamma(g)\Gamma(-\frac{1}{2}+b+e-\frac{n}{2})}{\Gamma(e)\Gamma(f)}\quad
_{4}F_{3}(b,1+b-d,-\frac{1}{2}+b+e-\frac{n}{2},\right.\nonumber\\
&&\left.-\frac{1}{2}+b+f-\frac{n}{2};1+b-c,\frac{1}{2}+b-\frac{n}{2},
-\frac{1}{2}+b+g-\frac{n}{2};1)+\Gamma(b-c)\right.\nonumber\\
&&\left.\times\frac{\Gamma(-1+c+e-\frac{n}{2})\Gamma(g)\Gamma(-1+c+f-\frac{n}{2})\Gamma(1-c+\frac{n}{2})}
{\Gamma(b)\Gamma(-c+d)\Gamma(-1+c+g-\frac{n}{2})\Gamma(e)\Gamma(f)}\quad_{4}F_{3}(c,1+c-d,\right.\nonumber\\
&&\left.-1+c+e-\frac{n}{2},-1+c+f-\frac{n}{2};1-b+c,c-\frac{n}{2},-1+c+g-\frac{n}{2};1)\right.\nonumber\\
&&\left.-\sqrt{-\frac{1}{p^{2}_{i}}}p_{i}\frac{\Gamma(b-c)\Gamma(-\frac{1}{2}+c+e-\frac{n}{2})\Gamma(g)\Gamma(-\frac{1}{2}+c+f-\frac{n}{2})\Gamma(\frac{1}{2}(1-2c+n))}
{\Gamma(b)\Gamma(-c+d)\Gamma(-\frac{1}{2}+c+g-\frac{n}{2})\Gamma(e)\Gamma(f)}\right.\nonumber\\
&&\left.\times_{4}F_{3}(c,1+c-d,-\frac{1}{2}+c+e-\frac{n}{2},-\frac{1}{2}+c+f-\frac{n}{2};1-b+c,\frac{1}{2}+c-\frac{n}{2},\right.\nonumber\\
&&\left.-\frac{1}{2}+c+g-\frac{n}{2};1)
-\sqrt{-\frac{1}{p^{2}_{i}}}p_{i}\frac{\Gamma(-\frac{1}{2}+b-\frac{n}{2})\Gamma(-\frac{1}{2}+c-\frac{n}{2})\Gamma(\frac{1}{2}+\frac{n}{2})}{\Gamma(-\frac{1}{2}+d-\frac{n}{2})\Gamma(c)\Gamma(b)}\right.\nonumber\\
&&\left.\times_{4}F_{3}(e,f,\frac{1+n}{2},\frac{3}{2}-d+\frac{n}{2};g,\frac{3}{2}-b+\frac{n}{2},\frac{3}{2}-c+\frac{n}{2};1)\right.\nonumber\\
&&\left.+\frac{\Gamma(-1+b-\frac{n}{2})\Gamma(-1+c-\frac{n}{2})\Gamma(1+\frac{n}{2})}{\Gamma(-1+d-\frac{n}{2})\Gamma(c)\Gamma(b)}\quad_{4}F_{3}(e,f,1+\frac{n}{2},2-d+\frac{n}{2};\right.\nonumber\\
&&\left.g,2-b+\frac{n}{2},2-c+\frac{n}{2};1)\right]
\end{eqnarray}
and integrals of the form
$\int_0^{\infty}\theta(p_{i}-p_{f})[_{2}F_{1}(b,c;d;\frac{p^{2}_{f}}{p^{2}_{i}})]^{2}
p^{\pm 2ik+n}_{f}dp_{f}$ which is the same type as the above and
have a closer result. The above integrals are valid in the
assumption $n\geq 0$.

\textbf{Acknowledgments}
\par
I would like to thank Professor Ion I.Cot\u aescu for reading the
manuscript and for useful suggestions and discussions that help me
to improve this work. Also I would like to thank Nistor
Nicolaevici for useful suggestions and to Racoceanu Radu for
helping me with numerical calculations.

\end{document}